\documentclass[pra,aps,superscriptaddress,twocolumn]{revtex4-2}

\usepackage{amsmath}
\usepackage{amssymb}
\usepackage{graphicx}
\usepackage[colorlinks=true,linkcolor=blue,citecolor=red,urlcolor=magenta]{hyperref}
\usepackage{xcolor}
\usepackage{bbold}
\newtheorem{theorem}{Theorem}
\newtheorem{proposition}[theorem]{Proposition}

\begin{document}

\title{How Likely Are You to Observe Non-locality with Imperfect Detection Efficiency and Random Measurement Settings?}

\begin{abstract}
    Imperfect detection efficiency remains one of the major obstacles in achieving loophole-free Bell tests over long distances. At the same time, the challenge of establishing a common reference frame for measurements becomes more pronounced as the separation between parties increases. In this work, we tackle both of these issues by examining the impact of limited detection efficiency on the probability of Bell inequality violation with Haar random measurement settings. We derive analytical lower bounds on the violation probability for a two-qubit maximally entangled state, which are tight for correlation inequalities and perfect detection efficiencies. We further investigate it numerically for more qubits and settings using two detection efficiency models and an original method based on the linear programming. Beyond that, we show that the so-called typicality of Bell inequality violation, i.e., almost certain violation of local realism with sufficiently many particles or random measurement directions, holds even if the detection efficiency is limited. Our findings reveal that increasing the number of measurement settings can compensate for efficiency losses above the critical threshold. Additionally, we determine critical detection efficiencies for both two-party and three-party scenarios. For two parties, we recover previously established results, while for three parties, we derive a symmetric critical efficiency of $\eta_{crit} = 2/3$ for the W and GHZ states within the binning model. In cases involving the no-detection outcome, we present a modified inequality using a pair of orthogonal observables for each party with $\eta_{crit} \approx 0.7208$, which is notably less than 0.75 for the GHZ state. These results offer deeper insights into the limitations and possibilities for certifying non-locality in the presence of limited detection efficiency, shedding light on its robustness in practical Bell tests.
    \end{abstract}

\author{Pawe{\l} Cie\'sli\'nski}
\email{pawel.cieslinski@ug.edu.pl}
\affiliation{Institute of Theoretical Physics and Astrophysics, University of Gda\'nsk, 80-308 Gda\'nsk, Poland}
\affiliation{International Centre for Theory of Quantum Technologies (ICTQT),
University of Gdansk, 80-308 Gda\'nsk,
Poland}

\author{Tam\'as V\'ertesi}
\affiliation{HUN-REN Institute for Nuclear Research, P.O. Box 51, H-4001 Debrecen, Hungary}

\author{Mateusz Kowalczyk}
\affiliation{Institute of Theoretical Physics and Astrophysics, University of Gda\'nsk, 80-308 Gda\'nsk, Poland}

\author{Wies{\l}aw Laskowski}
\affiliation{Institute of Theoretical Physics and Astrophysics, University of Gda\'nsk, 80-308 Gda\'nsk, Poland}

\maketitle

\section{Introduction}
Bell non-locality~\cite{Brunner_2014} is a well-established resource in quantum cryptography ~\cite{Ekert_1991,Mayers_1998, Barrett_2005, Acin_2007, Pironio_2009}, randomness generation~\cite{Pironio_2010,Colbeck_2011,Colbeck_2011_2, ArnonFriedman_2018} and communication ~\cite{Cleve_1997,Brukner_2002,Aolita_2012,Moreno_2020}. One of the main problems in taking full advantage of its capabilities is the imperfect detection efficiency~\cite{Pearle_1970,Larsson_1998,Pal_2012, Pal_2015,Cope_2019,Miklin_2022,Marton_2023}, which becomes especially daunting in the case of long-distance Bell experiments. On the other hand, a large separation between distinct parties complicates calibration and establishing a common reference frame for the measurement settings. Both of these problems are the key issues in performing quantum information protocols using satellites in space, see e.g. ~\cite{Bedington_2017, Sidhu_2021, Knips_2024}. Randomized measurements have proven themselves useful in many areas of quantum information~\cite{Cieslinski2024, Elben_2022}, among which is the possibility of certifying non-locality without the need for a shared reference frame~\cite{Cieslinski2024}. Thus, incorporating detection efficiency into the random measurements Bell test can address both of the mentioned problems.  
\begin{figure}[ht!]
    \centering
    \includegraphics[scale=0.72]{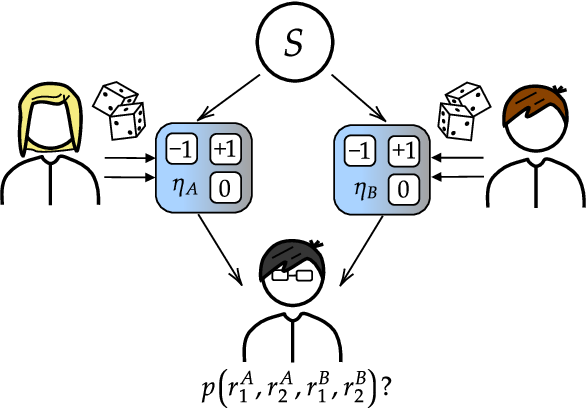}
    \caption{\textit{Probability of violation in a random measurements Bell test.} A quantum state is distributed between Alice and Bob. Using their imperfect apparatus with detection efficiencies $\eta_A$ and $\eta_B$ respectively they perform two randomly chosen measurements each. Once they gather the statistics, they send it to an external referee and ask if there exists a local realistic explanation for the experimental data. The probability of Bell inequality violation encompasses the average number of negative answers to this question. 
    }\label{fig:motivation}
\end{figure}

In this paper, we explore the dependence of the detection efficiency on the probability of local realism violation when the measurement directions are chosen at random. For simplicity, imagine a two-party scenario where Alice and Bob perform two Haar randomly chosen measurements (i.e. measurement directions chosen uniformly and randomly from a Bloch sphere) on a shared two-qubit state with imperfect detectors, see Fig.~\ref{fig:motivation}. Whenever the detectors click, they record $\pm 1$ and $0$ otherwise. Later we will also consider a different model for the imperfect detection efficiency based on the outcome binning strategy (also referred to as assignment strategy), however, for now, if not stated differently, we treat the no-detection event as $0$. Then, they gather their statistics and send it to an external referee who checks whether there exists a local realistic model, i.e,. a joint probability distribution compatible with the obtained information. How often will he be unable to do so? This is exactly the question of our interest.
Formally, one can define the probability of Bell inequality violation as~\cite{Barasinski_2020,Barasinski_2021}
\begin{equation}
    \mathcal{P}_V=\int d\Omega f(\Omega),
    \label{eq:probability_of_violation}
\end{equation}
where the integral goes over all the parameters within a Bell scenario, and $f$ is a binary function that outputs $1$ whenever the given settings allow one to violate some Bell inequality and $0$ otherwise. This distribution was already studied in the past in several ways and contexts, see e.g.~\cite{Liang_2010, Shadbolt_2012, Wallman_2012, Senel_2015, de_Rosier_2017, Lipinska_2018, Barasinski_2020, Yang_2020, de_Rosier_2020,  Pandit_2022, Barasinski_2021}. In practical calculations, the above integral is estimated by a large sampling ($\approx 10^9$ to ensure high significance of the numerical results) over the parameter space.
Moreover, this quantity can be treated as an equivalent description of non-locality for a full set of tight Bell inequalities in a given scenario~\cite{Barasinski_2019}. 

Here, we derive an analytical detection efficiency dependent lower bound on the probability of Bell inequality violation for a two-qubit maximally entangled state, which is exact for correlation inequalities and perfect efficiencies. We further explore this dependence numerically for more parties and settings using an original method based on linear programming first used in~\cite{de_Rosier_2017}. Our findings extend the previous results on the random measurements Bell tests by including the inevitable limitations of experimental devices and contain quantitative predictions for direct practical use. Since our approach does not refer to a specific inequality but encompasses an entire Bell scenario, we were able to determine the symmetric critical efficiencies $\eta_{crit}$ for the three-qubit GHZ as well as the W state and improve the best-known results for the two- and three-setting inequalities. In the case of perfect detectors, this problem was already studied in Refs.~\cite{Barasinski_2020, Liang_2010, de_Rosier_2017, Lipinska_2018}. 
One of the conclusions made in \cite{de_Rosier_2017, Lipinska_2018} was that with a growing number of particles or measurement settings, violation of local realism is almost certain - the so-called typicality of Bell violation. Our findings extend this conclusion to the case of imperfect detectors and furthermore show that any drop in efficiency above the critical one can be compensated with additional settings. This shows that the misalignment of the measurement directions and a drop in detection efficiency can be overcome with a higher number of measurements.

\section{Lower bound on the probability of violation for a two-qubit maximally entangled state}\label{sec:lower_bound}

Consider a probability-based CHSH inequality
\begin{equation}
    I=\sum_{x,y=0}^1 P(a \oplus b =xy|x,y) \leq 3,
    \label{eq:chsh}
\end{equation}
violated maximally by $| \psi^- \rangle=(|01\rangle - |10 \rangle)/\sqrt{2}$ with suitably chosen measurement directions.
By including detection efficiencies $\eta_A$ and $\eta_B$ as well as the optimal local strategies, it can be expressed as
\begin{eqnarray}
    I(\eta_A, \eta_B)&=&\eta_A (1-\eta_B) M_A+\eta_B (1-\eta_A) M_B \\
    &+& \eta_A \eta_B Q + (1-\eta_A)(1-\eta_B) X \nonumber\\  
    &=& 2 \eta_A (1-\eta_B)+2 \eta_B (1-\eta_A) \nonumber \\
    &+& \eta_A \eta_B Q + 3 (1-\eta_A)(1-\eta_B) \leq 3 \nonumber
    \label{eq:chsh_eta}
\end{eqnarray}
where $M_i$ and $X$ denote the CHSH value when either one or both detectors did not click respectively, and $Q$ corresponds to $I$~\cite{Marton_2023}. The main premise of this work is to determine the probability of observing  $I(\eta_A,\eta_B)>L=3$ when the measurements are chosen randomly according to the Haar measure~\cite{Cieslinski2024}. Examining a single CHSH variant is sufficient to determine the probability of interest as the two-setting two-output scenario is fully characterised by a set of four CHSH inequalities with each of them having an equal contribution to $\mathcal{P}_V$~\cite{Liang_2010}.
Note that, whenever $\Tilde{Q}=\eta_A \eta_B Q + 3 (1-\eta_A)(1-\eta_B) > L$, the inequality has to be violated. Thus, determining the probability of $\Tilde{Q}$ violation denoted as $\mathcal{P}_V^c$ would yield a lower bound on $\mathcal{P}_V$ as a function of the detection efficiencies. Note that such an approach will be exact for perfect $\eta$ and correlation-based inequalities. We also expect that it should approximate the exact probability up to a small percentage error in some range of large $\eta$. 
Following this idea, we derive the considered bound $\mathcal{P}_V^c$ by extending the proof presented in ~\cite{Liang_2010} to our scenario. All details are included in Appendix~\ref{app:proof} and lead to the following:
\begin{proposition}
For the asymmetric detection efficiency $\eta_A=1, \eta_B=\eta$ the probability of violation $\mathcal{P}_V$ is lower bounded by
\begin{eqnarray}\label{boundeqn}
   & \mathcal{P}_V^c(\eta)=\frac{1}{\eta ^2 \left(\eta ^2-1\right)}\bigg[ \pi + 2 (\eta^2-1) \kappa - \pi \eta^4  \\
   &+ 4 \sqrt{2}  \left(\sqrt{\eta^2 + \kappa} + \sqrt{\kappa^2 (\eta^2 + \kappa)} - 2  \eta^2  \sqrt{\eta^2 + \kappa}\, \right) \nonumber \\
   & + 8 \left(\eta ^4-1\right) \mathrm{arccot} \left \lbrace \sqrt{2} \sqrt{\frac{1}{\frac{\kappa}{\eta ^2}+1}}+\sqrt{\frac{\eta ^2-\kappa}{\eta ^2+\kappa}}\right \rbrace \bigg]; \quad \eta >\frac{1}{\sqrt{2}} \nonumber,
   \label{eq:bound}
\end{eqnarray}
where $\kappa=\sqrt{2 \eta ^2-1}$. 
\end{proposition}

\begin{proposition}
For the symmetric detection efficiency $\eta_A=\eta_B=\eta$ the probability of violation $\mathcal{P}_V$ is lower bounded by
\begin{eqnarray}\label{boundeqns}
    &\mathcal{P}_V^c(\eta)=\frac{2}{\eta^4} \bigg[2 \gamma + \eta ^2\pi [(\eta -2) \eta +2] \\
&+ \eta(\eta -2) \bigg[2 \sqrt{2} \big(\sqrt{\gamma  \eta ^2+2 \gamma  (1-\eta )+\eta ^4}\nonumber\\
&- \sqrt{-\gamma  \eta ^2-2 \gamma  (1-\eta )+\eta ^4}\big)+\gamma \bigg]+2 \gamma \nonumber\\
& -4 \eta ^2 [(\eta -2) \eta +2] \arctan \left \lbrace \frac{4 \sqrt{(\eta -1)^2 ((\eta -1) \eta +1)^2}}{\gamma  ((\eta -2) \eta +2)+\eta ^4}\right \rbrace \bigg];\nonumber\\
&\eta >  \sqrt{5 + 4 \sqrt{2}}- \sqrt{2} -1 \nonumber,
   \label{eq:bounds}
\end{eqnarray}
where $\gamma = \sqrt{ 8( \eta - \eta^2) + 4 \eta^3 + \eta^4-4}$. 
\end{proposition}
Constraints on the allowed values of $\eta$ reflect the upper bounds on the critical efficiencies. The above bounds agree well with the linear programming results and can recover $\mathcal{P}_V$ with good accuracy in an entire range of $\eta$. For a direct comparison with numerics, see Sec.~\ref{sec:numerics}.

\section{Numerical results}
\label{sec:numerics}
An alternative approach to this problem, in general, is to search for the joint probability distributions of the form
\begin{equation}
    P_{LR}(r^1_1, r^1_2,\cdots, r^1_{m^1}, r^2_1, \cdots, r^N_{m^N})=P_{LR}(\Vec{r}^{\,1}, \cdots, \Vec{r}^N) \nonumber,
\end{equation}
describing the statistics of measurement outcomes $r^{\mathcal{O}}_k$ obtained for the party $\mathcal{O}$ measuring the setting $k$. For example, in the two-party, two-setting scenario for qubits, we have $P_{LR}(r^1_1, r^1_2, r^2_1, r^2_2)=P_{LR}(r^A_1, r^A_2, r^B_1, r^B_2)$, where $A$ and $B$ stand for Alice and Bob, respectively, and $r^{\mathcal{O}}_k=\pm 1$. For the sake of generality, we will introduce all of the concepts using number labelling of the parties. Later, however, we will constrain ourselves to at most three parties and use alphabetical labels for clarity. In general, quantum mechanics allow one to measure probability distributions that do not adhere to the above. Let $M^{\mathcal{O}}_k=\sum_{r^i=\pm 1} r^i_k |r^i_k \rangle \langle r^i_k|$ be the $k$th observable measured by $\mathcal{O}$. For the given state $\rho$ quantum mechanical probabilities yield
\begin{eqnarray}
    & P_Q(r^1_{k_1},  \cdots, r^N_{k_N}|M^1_{k_1}, \cdots, M^N_{k_N}) \\
    & =\mathrm{Tr} \left[\rho |r^1_{k_1} \rangle \langle r^1_{k_1}| \otimes \cdots \otimes |r^N_{k_N} \rangle \langle r^N_{k_N}| \right]. \nonumber
\end{eqnarray}
If the local realistic model for this distribution exists, then the above probabilities can be explained by the following marginal distributions
\begin{eqnarray}
    & P_Q(r^1_{k_1},  \cdots, r^N_{k_N}|M^1_{k_1},\cdots, M^N_{k_N}) \\
    & \overset{\text{\tiny{LR}}}{=}\displaystyle \sum_{r^1_{k'_{1}}, \ldots , r^N_{k'_{N}}=0}^{d-1} P_{LR}(\Vec{r}^{\,1}, \cdots, \Vec{r}^N)=P_{LR}(r^1_{k_1},  \cdots, r^N_{k_N}), \nonumber
\end{eqnarray}
where $k_i'\neq k_i (i=1,\cdots, N)$. The above represents the probability of obtaining measurement outcomes $r^1_{k_1},  \cdots, r^N_{k_N}$ when measuring observables $M^1_{k_1},\cdots, M^N_{k_N}$.  Again, in the two-party two-setting scenario for qubits we have, e.g. $P_{LR}(r^1_1, r^2_1)=P(r^1_1,-1, r^2_1,-1)+P(r^1_1,-1, r^2_1,1)+P(r^1_1,1, r^2_1,-1)+P(r^1_1,1, r^2_1,1)$. 
To include the imperfect detection efficiencies we treat the local events where no particle was detected as an additional measurement outcome 0. As a result, more probabilities $P_{LR}$, with $r_k^{\mathcal{O}}=0,\pm1$, that need to be taken into account~\cite{Durt_2001}.
Finally, if the measurement statistics is incompatible with the joint probability distribution description, then no local realistic model exists, and thus, there is a Bell inequality that is violated with the given state and settings~\cite{Fine_1982, Kaszlikowski_2000}.  It is worth noting that such inequality can be violated in a very weak manner~\cite{de_Rosier_2020}. This strongly points out the relevance of including the detection efficiency in such a Bell scenario.
For completeness we will also consider a model in which the outcomes are grouped in just two categories. There, we will consider a single imperfect detector per party that records the $+1$ events. If no $+1$ was detected, we treat it as $-1$. However, if not stated differently, we will work with the first scenario.

Before we proceed to the numerical results, we will specify the notion of random measurements in the studied context. Randomised settings are performed by applying a unitary transformation from the Haar ensemble $U$ on the projection operators associated with the Pauli matrix $\sigma_z$, i.e. $U_{[k]} |i \rangle \langle i | =  |h^i_k \rangle \langle h^i_k|$, $i=0,1$. The explicit form of $U_{[k]}$ as a function of measurement angles can be found in \cite{de_Rosier_2017}. Using such parametrisation, Haar sampling reduces to drawing angles from a uniform distribution. Generating $m^{\mathcal{O}}$ unitaries will produce $m^{\mathcal{O}}$ random settings, enumerated by $k$, for party $\mathcal{O}$. For each sample of tensor product of Haar unitaries $U_{[k_i^{1}]}\otimes \cdots \otimes U_{[k_i^{N}]}$ and a given state, we generate the measurement statistics and look for a local realistic model as presented above using the linear programming method first used in~\cite{de_Rosier_2017}. If the joint probability distribution does not exist for the produced statistics we put $f=1$ ($f=0$ otherwise), and by having a large enough sample, estimate $\mathcal{P}_V$ (\ref{eq:probability_of_violation}) which is the quantity of our interest. For all of our calculations, the number of random unitaries varies between  $10^3-10^{10}$ depending on the number of studied settings and parties.

\begin{table}[]
\centering
\caption{Summary of the symmetric critical detection efficiencies for the studied states. The first column of the table refers to the state under consideration, and is followed by the imperfect detection model and the number of settings $m$ in the Bell scenario. The subscripts in the state's name refer to the number of parties. Note that beyond recovering the previously known results for two-qubits, we also present new critical detection efficiencies in the tripartite scenario. For more details, see the main text.}
\label{tab:crit}
\begin{tabular}{c|c|c|c}
state & detection model & $m$ & $\eta_{crit}$   \\ \hline
$|\psi^-\rangle $ & both & 2 &  $2/(1+\sqrt{2})\approx 0.8284$ \\
$|GHZ_3 \rangle$ & three-outcome & 2 & $(\sqrt{10}-1)/3\approx 0.7208$    \\
$|GHZ_3 \rangle$ & binning & 2,3 & $2/3$   \\
$|W_3 \rangle$ & binning & 2,3 &  $2/3$ 
\end{tabular}
\end{table}

\subsection{Two-qubit maximally entangled states}
\label{sec:two-qubits}
\begin{figure}
    \centering
    \includegraphics[width=0.45 \textwidth]{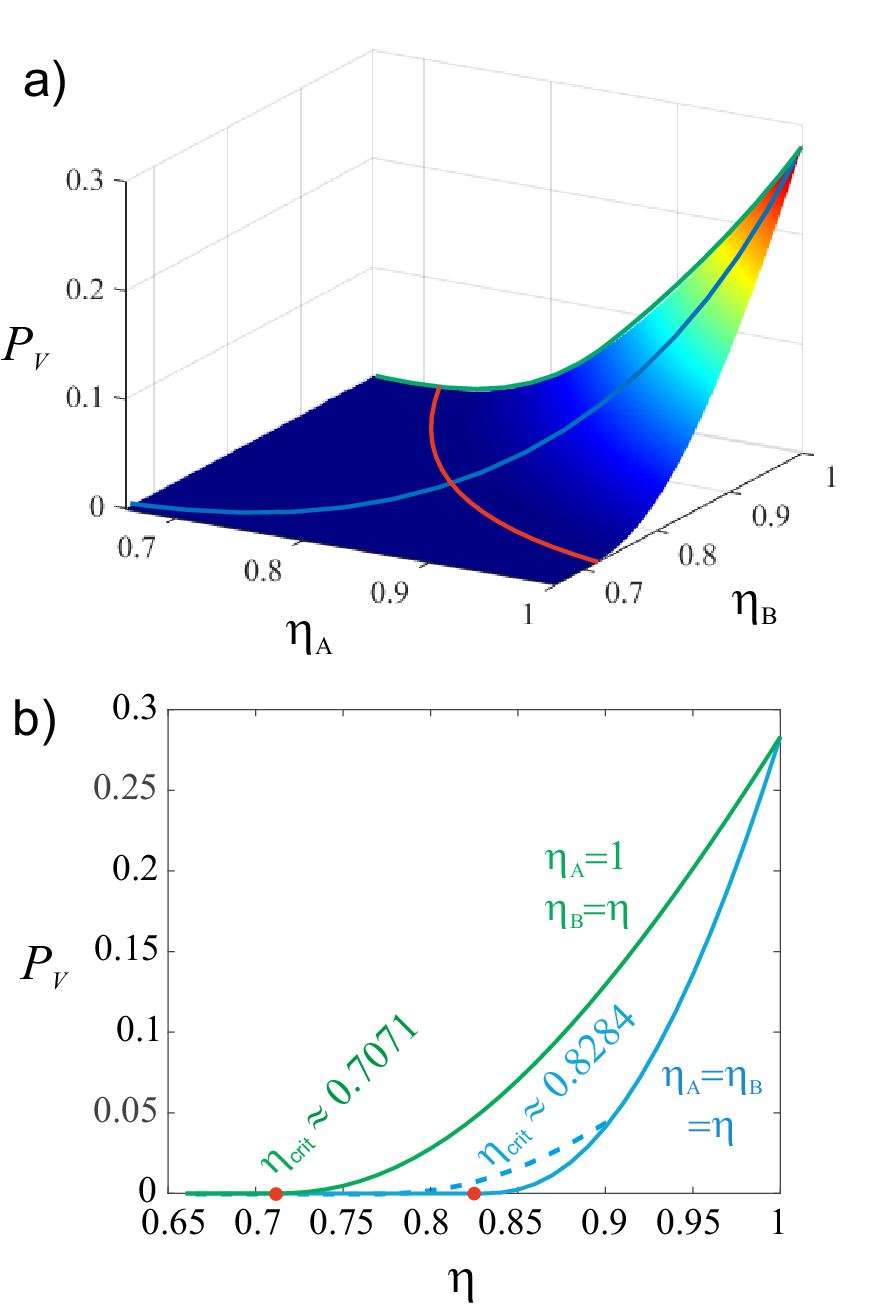}
    \caption{Probability of Bell inequality violation $\mathcal{P}_V$ as a function of detection efficiencies $\eta_A$ and $\eta_B$ for a two-qubit maximally entangled state and two settings per party. Each point on the plot was computed with a sample of $10^{6}-10^{10}$ linear programming evaluations and a step of $\Delta \eta=10^{-3}$. The solid red line on the $a)$ panel represents the critical efficiency $\eta_{crit}$ and separates the region of $\mathcal{P}_V$ being $\approx 0$ up to statistical significance and numerical error. The additional coloured lines correspond to the two significant cases of symmetric $(\eta_A=\eta_B=\eta)$ and asymmetric $(\eta_A=1, \eta_B=\eta)$ detection efficiency scenarios marked in blue and green respectively. For clarity, they are plotted separately on panel $b)$. There, an additional case of $\mu_A=0.9$ and $\mu_B=\mu$ was plotted (blue dashed curve) to illustrate a more nuanced asymmetric scenario.
    As expected, the maximal $\mathcal{P}_V$ occurs for the perfect detection case and reproduces the known result derived in Ref.~\cite{Liang_2010}. As the efficiencies drop down, the probability of violation decreases monotonously, and here we provide the exact predictions for its value at any given $(\eta_A,\eta_B)$. The above numerical results obtained through the linear programming are in good agreement with the derived lower bounds (\ref{eq:bound}) and (\ref{eq:bounds}). For more details, see the main text.}
    \label{fig:GHZ2_results}
   \end{figure}

Results for a two-qubit maximally entangled state and two settings per party obtained through the discussed techniques are shown in Fig.~\ref{fig:GHZ2_results}. The general behaviour of the probability of violation agrees with a basic intuition as it decreases monotonously with each of the efficiencies $\eta$. However, here the important information lies not only in the general behaviour of the probability of interest but in our ability to make quantitative predictions on $\mathcal{P}_V$. The range of the studied efficiencies goes from the approximate $\mu_{crit}$ to 1. Note that it covers $\mu$ which allow for the device independent quantum key distribution, see e.g.~\cite{Xu2022} that tolerates efficiencies above $\approx 0.685$, and its experimental demonstration~\cite{Liu2022}. For the perfect detection case, we recover $\mathcal{P}_V \approx 28.32\%$~\cite{Liang_2010, de_Rosier_2017}. On the other hand, we can localise the critical efficiencies yielding $\mathcal{P}_V \approx 0$, up to statistical significance and numerical error, for any pair of $(\eta_A, \eta_B)$ denoted by the red line. Their values in the symmetric (blue line) and asymmetric case (green line) coincide with the known results~\cite{Eberhard_1993,Gisin_1999}. Note that in many important situations where the entangled states are distributed between long-range separated parties, the detection efficiency is drastically reduced due to the transmission losses. In a direct scenario, loophole free violation of any standard Bell inequality is impossible. However, modern quantum repeaters or memories promise a solution to this problem and can effectively pass most of the detection efficiency to the measuring devices~\cite{Sidhu_2021}. Assuming that, take an example of space-to-ground links where the asymmetric detection efficiency scenario can be better described when $\mu_A<1$~\cite{Marsili2013, deForgesdeParny2023}, e.g. $\mu_A=0.9$ and $\mu_B=\mu$. In this case, the probability of violation drops as compared to the standard asymmetric scenario and is plotted explicitly with a dashed line in Fig.~\ref{fig:GHZ2_results}.

We further observe that in the asymmetric case the lower bound $\mathcal{P}_V^c$ (\ref{eq:bound}) agrees with $P_V$ up to $\approx 3.5\%$ relative error within the whole range of $\eta$, i.e. $\eta_{asym} \in [\eta_{crit}^c,1]$, where $\eta_{crit}^c=1/\sqrt{2}$. In the symmetric case (\ref{eq:bounds}), the same holds up to $\approx 2.2\%$ error.

\subsubsection{Binning results for \texorpdfstring{$N=2$}{TEXT}}

The problem of $\mathcal{P}_V$ computation can also be approached with the binning model for imperfect detection efficiency. In such a scenario, there only exist two distinct outcomes associated with a two-element positive operator valued measure (POVM)
\begin{eqnarray}
    P^{\mathcal{O}}_+&=&\eta_{\mathcal{O}} |h^0_k\rangle \langle h^0_k| \\
    P^{\mathcal{O}}_-&=& \mathbb{1} - P^{\mathcal{O}}_+, \nonumber
\end{eqnarray}
where again $|h^0_k\rangle \langle h^0_k|$ is a random projective measurement associated with the $+1$ outcome. Results for the symmetric and asymmetric scenarios obtained within that framework are practically indistinguishable from the ones presented in the previous section.  
This is due to the fact that within the limited detection efficiency two-setting two-party scenario, the three-outcome approach does not provide any extra inequalities~\cite{Branciard_2011}.
However, this will no longer be the case for the three-qubit states in the subsequent section.

\subsection{Three-qubits W and GHZ states}
\label{sec:three-qubits}

\begin{figure}
    \centering
    \includegraphics[width=0.43\textwidth]{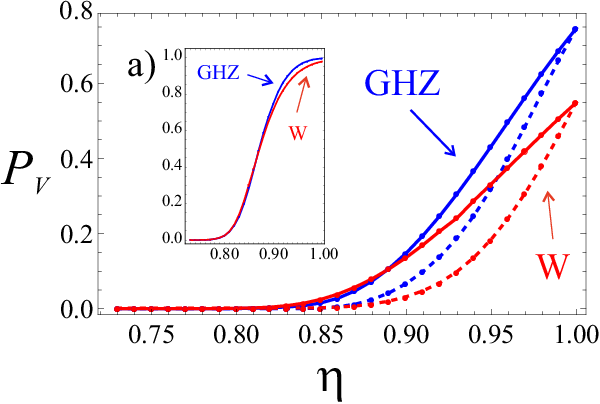}
    \caption{Probability of violation for the three-qubit GHZ (blue) and W (red) state with symmetric detection efficiencies $\eta_A=\eta_B=\eta_C=\eta$ and an extra no-detection outcome (solid line) and binning (dashed line for) models. 
    For small efficiencies, in the three-outcome model, the probability of violation is higher for the W state in both $m=2$ and $m=3$, see inset $a)$, cases. Then, both curves cross. For the $m=2$ case, it happens at $\eta = 0.89$ and for $m=3$ it can be localised between $\eta = 0.86$ and $\eta = 0.87$. Afterwards, the GHZ state probability grows faster and reaches its maximal value at $\eta=1$, which differs from the W state maxima significantly for $m=2$. The highest $\mathcal{P}_V$ is observed for the GHZ state and $m=3$ scenario, where it reaches almost $100\%$. Note that for the extra no-detection outcome model and both states, the probability of Bell inequality violation for $m=3$ as compared to $m=2$ is greater for all $\eta$. This shows that introducing more settings can increase the probability of violation in the imperfect detection randomised scenarios. In general, both states give similar predictions for the overall behaviour of $\mathcal{P}_V$, and this also holds for the binning model. Nevertheless, in a quantitative sense, the binning model performs worse and the violation probability is always smaller.  
    }
    \label{fig:three_qubits}
\end{figure}

For three qubits, we focus on two entangled states which are inequivalent under local unitaries and can yield different probabilities in the randomised scenario, i.e., the GHZ and W states defined as
\begin{eqnarray}
    |GHZ \rangle = \frac{1}{\sqrt{2}}\left(|000\rangle + |111\rangle \right), \\
    |W\rangle=\frac{1}{\sqrt{3}}\left(|100\rangle + |010\rangle + |001\rangle \right).
\end{eqnarray}
Here, we will examine two- and three-setting Bell inequalities in the same context. By performing analogous calculations, we evaluate the probability of violation in the case of symmetric efficiencies $\eta_A=\eta_B=\eta_C=\eta$ and a step of $10^{-2}$, see Fig.~\ref{fig:three_qubits}. We observe that for small efficiencies, the probability of violation is higher in the case of the W state for both $m=2$ and $m=3$ measurement settings. Then, both curves cross. It is interesting from the non-local resources point of view due to the mentioned inequivalence of the studied states. From a practical perspective, if we know our efficiencies, it turns out that it might be as beneficial to prepare the W state as the GHZ one depending on their generation feasibility. After this point, the GHZ state probability exceeds the one for the W states and reaches its maximum value for either choice of $m$. Interestingly, the probability of Bell inequality violation for $m=3$ is greater for all $\eta$ as compared with the two-setting case and the maximum $\mathcal{P}_V$ for both states is close to 1 (see inset $a)$ in Fig.~\ref{fig:three_qubits}). This points to the fact that introducing more settings increases the probability of violation, thus expanding the previous results~\cite{de_Rosier_2017,Lipinska_2018} to the imperfect detection scenarios. This observation will be explored in more detail in the subsequent section. 

For the three-qubit case, we can also approximate the critical detection efficiencies $\eta_{crit}$ without referring to a specific inequality. However, due to its statistical nature, it has an intrinsic statistical and numerical uncertainty. 
Up to our knowledge, the best previous results for $\eta_{crit}$ and the studied states are $\eta_{crit}(GHZ_3)=0.75$~\cite{Larsson_1998} for $m=2$ and can be reduced only by introducing many ($m=17$) additional settings~\cite{Pal_2012}. For the W state they are given as  $\eta_{crit}(W_3)=0.83747$ for $m=2$ and $\eta_{crit}(W_3)=0.6$ for $m=3$~\cite{Pal_2015}.

From our data, we observe that the approximated $\eta_{crit}$ are smaller than the previously known best results. In order to determine them in a reliable manner, one would have to find the corresponding inequalities and prove that the violation vanishes for a given $\eta$. Unfortunately, this is a very daunting task in such a model. Nevertheless, we have found an inequality that improves the Mermin inequality efficiency threshold in such an approach. See Appendix~\ref{app:ghz} for details. By taking into account the non-signalling constraints and rewriting the original problem in the Collins-Gisin format, we arrived at the inequality with $\eta_{\text{crit}}=(\sqrt{10}-1)/3\approx 0.7208$. The quantum value was evaluated on the rotated GHZ state, but since it is equivalent to the original state under local unitaries they are indistinguishable in the random measurement setting, and our reasoning is consistent. 

Despite finding an inequality that allowed us to lower the $\eta_{crit}$, we found that this more complex approach
is not necessary. By turning off one of the detectors, here the one measuring the $-1$ events, for each of the observers and moving onto the binning model, we observe that even smaller $\eta_{crit}$ can be found. Now, we will follow this path.

\subsubsection{Binning results for \texorpdfstring{$N=3$}{TEXT} and critical detection efficiencies}

Again, we shift our focus to the binning model. As above, we will limit ourselves to the symmetric detection efficiency scenario parametrised by a single $\eta$. The numerical results for the $\mathcal{P}_V$ are presented in Fig.~\ref{fig:three_qubits}. Here we see strong quantitative differences, especially for the W state. The probabilities in both approaches to the detection efficiency are, as expected, equal for $\eta=1$, but then drop down faster than before. The general behaviour stays the same. 

The advantage of such a formulation of the detection efficiency model is that the observables are standard qubit POVMs, and it becomes feasible to find the exact inequalities violated for small $\eta$ and derive the critical detection efficiencies. From the numerical calculations, we find the measurement settings for which only few joint probability distributions were found and $\mathcal{P}_V \approx 0$. Then we use the linear programming method introduced in \cite{Cieslinski_2024} to find the corresponding inequalities. For both of the studied states and $m=2$ we arrive at $\langle I_{C}\rangle \leq 1$ with
\begin{equation}
    I_{C}=-A_1 +\frac{1}{2}(\mathbb{1}+A_1)[B_1C_1+B_2C_1+B_1C_2-B_2C_2] \label{eq:I_C}
\end{equation}
where $X_i$ for $i=1,2$ and $X=A,B,C$ are the observables for each party. Note that there is only one non-trivial setting on Alice's side. The term $1/2(\mathbb{1}+A_1)$ can be interpreted as a projection and is known to improve the detection efficiency, see e.g. \cite{Kostrzewa_2018}. Note that the remaining part of $I_{C}$ is the CHSH expression, which immediately points to the previous work on a similar subject, e.g. ~\cite{Eberhard_1993, Cavalcanti_2011, Cavalcanti_2012, Chaves_2014}. Moreover, this inequality is tight as it is a facet of the Bell-Pitowsky polytope~\cite{Sliwa_2003}. Now, solving $I_C$ for $\eta$ and maximizing the violation we get $\eta_{crit}=2/3$. This is exactly the minimal detection efficiency for two-qubit states obtained in \cite{Eberhard_1993}. Moreover, after a suitable projection, the state on which the CHSH expression is measured is the Eberhard state
\begin{equation}
    |\psi_E\rangle=\frac{1}{\sqrt{1+\alpha^2}} \left( |00\rangle + \alpha |11\rangle  \right), 
\end{equation}
for some $\alpha \ll 1$. From the experimental perspective, the observed violations have a clear filtering interpretation. Performing analogous optimisation for $m=3$ we again arrive at (\ref{eq:I_C}) and $\eta_{crit}=2/3$.

All of the obtained symmetric critical detection efficiencies are summarized and presented in Table~\ref{tab:crit}.

\section{Typicality of local realism violation and efficiency compensation}
\label{sec:typicality}
\begin{figure}
    \centering
    \includegraphics[width=0.45\textwidth]{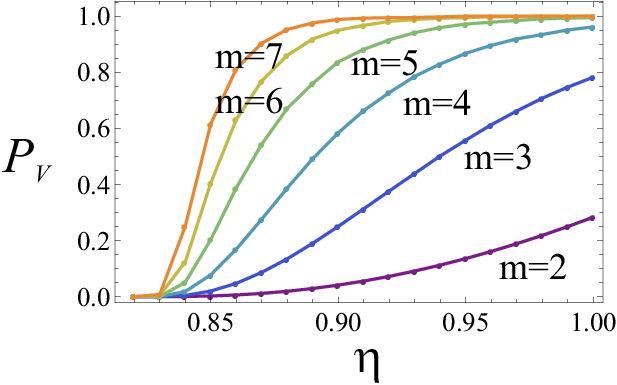}
    \caption{Probability of Bell inequality violation $\mathcal{P}_V$ as a function of symmetric detection efficiency $\eta$ for a two-qubit maximally entangled state and various number of settings $m$. Each colour represents a different $m \in [2,7]$. For $m=2$ settings, the probability grows slowly with $\eta$.  For $m \geq 3$ its behavior can be described through a sigmoid-type curve. Its break point moves toward lower efficiencies, and the slope increases with $m$, causing the $\mathcal{P}_V$ to grow exponentially past the critical efficiency and then saturate to almost $100\%$ for $m>4$. The above results confirm the typicality of non-locality in the imperfect detection scenario and show how a drop in detection efficiency can be compensated with an additional setting.}
    \label{fig:many_settings}
\end{figure}

\begin{figure}[h!]
    \centering
    \includegraphics[width=0.43\textwidth]{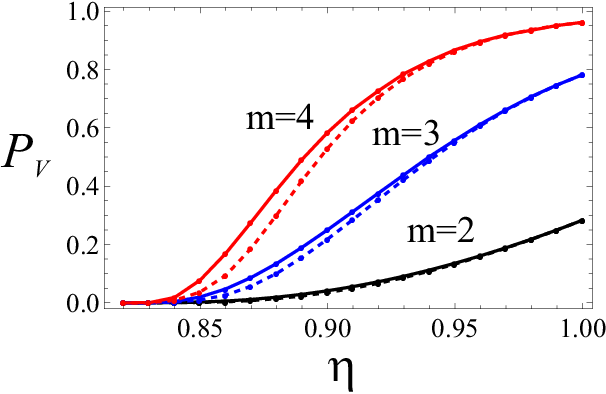}
    \caption{Probability of Bell inequality violation $\mathcal{P}_V$ for a two-qubit maximally entangled state and different models of imperfect detection efficiency in the symmetric scenario with $m$ settings per party. Coloured points connected by solid lines correspond to the scenario where the no-detection event is treated as an extra outcome. Predictions of the binning model are plotted with dashed lines. In spite of the number of settings $m$, both models give similar results for efficiencies close to 1. Then, for smaller $\eta$ they start to diverge and differ stronger the greater the $m$. In the scenario with $m=2$, they overlap as the three-outcome approach does not provide any extra inequalities in this case~\cite{Branciard_2011}. In all the cases, the no-detection extra outcome model yields a greater (or equal) probability of violation.}
    \label{fig:comparison}
\end{figure}

In Ref.~\cite{de_Rosier_2017,Lipinska_2018} the authors showed that with perfect detectors, non-locality is typical, meaning that by extending the number of particles or randomly chosen settings the local realism violation becomes almost certain, i.e. $\mathcal{P}_V \approx 100\%$. Here, we investigate this problem in a more practical scenario. We determined the probability of Bell inequality violation for the maximally entangled two-qubit state and the whole range of symmetric detection efficiencies with a step of $10^{-2}$ and different number of measurement settings for Alice and Bob. For clarity, we chose $m^A=m^B=m$ and performed the analysis for $m \leq 7$. The results are depicted in Fig.~\ref{fig:many_settings}. The differences in predictions for the extra no-detection outcome and the binning models for $m <5$ are shown and discussed in Fig.~\ref{fig:comparison}. The obtained results show that increasing the number of settings significantly enhances the probability of violation. For instance, with $m=2$ settings, the probability grows slowly and reaches the previously discussed maxima, whereas for more than two settings a rapidly growing sigmoid-type curve appears from the data. The breakpoint moves toward lower $\eta$'s, and the slope increases with the number of added settings, causing the $\mathcal{P}_V$ to grow exponentially past the critical efficiency for higher $m$'s. For $m>4$ the probability of violation saturates to $\approx 100 \%$ and this process occurs faster the greater the $m$. In the limit of $m \rightarrow \infty$, the expected behaviour follows a step function past the critical efficiency that immediately reaches a certain violation~\cite{Lipinska_2018}. This is intuitively due to the certainty of sampling the optimal CHSH measurement angles. We thus conclude that typicality is preserved even when our detection efficiency is limited. Adding new measurement settings causes $\mathcal{P}_V$ to eventually reach $\approx 100\%$. Such asymptotic behavior can be analyzed in terms of statistical significance through, for example, the Wilson technique~\cite{Wilson1927}. Given the confidence level of $95\%$ and the observed $\mathcal{P}_V=100\%$, the lower bound on the estimated probability is $99\%$ with $\approx 270$ events and $99.9\%$ with $\approx 2700$ events. For a confidence level of $99\%$ it changes to $\approx 460$ and $\approx 4600$ respectively. Clearly, a few thousand samples are enough to infer the $\mathcal{P}_V> 99\%$.

An interesting observation from this data is that our detection capabilities can be compensated by performing more measurements. Naturally, this holds as long as we are past the critical efficiency below which the probability is always 0. This observation is consistent with the previous results on Bell non-locality showing the advantage of multi-setting Bell inequalities, see e.g.~\cite{Laskowski2004, Son2006, Nagata2006, Vertesi2010, Pal_2012, Pal2015b, Bae2018}. It is worth noting that the observed behaviour of the probability of interest is similar to the one describing $\mathcal{P}_V$ as a function of visibility for a two-qubit Werner state~\cite{Shadbolt_2012}, which should follow the symmetric $\mathcal{P}_V^c$ (\ref{eq:bound}) in the $m=2$ case. Moreover, our results corresponding to the perfect detection efficiency are in agreement with the perfect visibility case. At last, note that for $m = 4$ there exists a tight Bell inequality that allows for a slightly lower detection efficiency of $\eta_{crit} = 0.8214$ for the two-qubit maximally entangled state, compared to the CHSH critical efficiency of $\eta_{\text{crit}} = 0.8284$~\cite{Brunner_2008}.

We further investigate the observation of efficiency compensation by calculating the minimal number of settings needed for a given detection efficiency to increase $\mathcal{P}_V$ up to the maximal value of $\approx 28.32\%$ from the two-setting perfect detection case. 
When increasing them from 2 to 3, the first value of $\eta$ that allows one to achieve such a task is $\eta = 0.91$, at which $\mathcal{P}_{V}=31.31 \%$. For $m = 4$, the required efficiency decreases to $\eta = 0.88$ with the corresponding probability of $38.71 \%$. Similarly, for $m=5$ we have $\eta = 0.86$ with the probability of $38.78 \%$. Finally, for $m=6$, the efficiency needed to outperform the perfect detection scenario with two settings is $\eta = 0.85$ with the probability of violation of $40.55\%$. The next question we ask is, when does the $\mathcal{P}_V$ increase the most while increasing the number of settings. For this, we calculate the relative increase in the probability of violation for different choices of the number of settings as $(\mathcal{P}_{V_{m_1}} - \mathcal{P}_{V_{m_2}}) / (\mathcal{P}_{V_{m_1}})=\Delta \mathcal{P}_{m_1 \rightarrow m_2}$, where $\mathcal{P}_{V_{m_1}}$ and $\mathcal{P}_{V_{m_2}}$ are the violation probabilities for the fixed number of settings $m_i$.
The acquired results are presented in  Table~\ref{tab:prob_change} in Appendix~\ref{app:table}. The greatest relative jump appears when one changes the number of settings from $2 \rightarrow 3$ with a probability growth factor of up to $1404.25\%$ for $\eta= 0.83$. Similarly, the transition from $3 \rightarrow 4$ shows initial growth of $429.41 \%$ at $\eta=0.83$, which then drops to $22.95 \%$ at $\eta = 1$. 
These results highlight the benefit of adding more settings to our test when the detection efficiency available drops down.

\section{Conclusions}

The probability of Bell inequality violation can be heavily influenced by the detection efficiency of our set-up. In this paper, we have exploited this dependence in two detection efficiency models thoroughly by determining its values for several states and varying number of measurement settings. Our results contain analytical lower bounds as well as extensive numerical calculations based on linear programming. We have confirmed the basic intuition behind its behavior and made new interesting observations. Any drop in detection efficiency for the two-qubit maximally entangled state above the critical one can be compensated by adding new random measurement settings to our Bell test. Preliminary data for the three-qubit GHZ and W states suggest that this behavior extends to a multipartite scenario. Based on the same calculations, we confirmed the existence of typicality of non-locality in the case of imperfect detection efficiency by reporting that introducing additional random settings to the set-up causes the probability of violation to eventually reach $\approx 100 \%$. Furthermore, we took advantage of the generality of random measurements scenarios and determined the critical efficiencies $\eta_{crit}$ for the three-qubit GHZ and W state, improving the previously best-known results.

These findings represent a significant step forward in understanding the practical limitations and possibilities of certifying non-locality, particularly in the context of long-distance Bell tests, such as those conducted via satellite-based communication. The obtained results and observations contain quantitative predictions that can help to improve current days and forthcoming experiments. Future work in this subject could focus on the higher number of particles, different states (including higher dimensions) for which the critical efficiency is unknown and genuine or network scenarios.

\section*{Acknowledgements}

We thank Marcin Paw{\l}owski for helpful discussions. 
PC and WL are supported by the National Science Centre (NCN, Poland) within the OPUS project (Grant No. 2024/53/B/ST2/04103) and MK within the Chist-ERA project (Grant No. 2023/05/Y/ST2/00005). PC acknowledges the support of Foundation for Polish Science (FNP) within the START programme. TV acknowledges the support of the EU (CHIST-ERA MoDIC), the National Research, Development and Innovation Office NKFIH (No. 2023-1.2.1-ERA\_NET-2023-00009 and No. K145927), and the ``Frontline'' Research Excellence Program of the NKFIH (No. KKP133827). This work is partially carried out under IRA Programme, project no. FENG.02.01-IP.05-0006/23, financed by the FENG program 2021-2027, Priority FENG.02, Measure FENG.02.01., with the support of the FNP. 
For the purpose of Open Access, the authors have applied a CC-BY public copyright licence to any Author Accepted Manuscript version arising from this submission. All data supporting the figures presented in the paper are available in the arXiv source files at \href{https://doi.org/10.48550/arXiv.2503.21519}{arXiv:2503.21519}.

\appendix

\section{Derivation of the lower bounds on the probability of violation}
\label{app:proof}
Here, we will derive the lower bound on the probability of local-realism violation $\mathcal{P}_V^c(\eta_A, \eta_B)$ presented in the main text. Since all maximally entangled states of two qubits are equivalent under local unitary transformations and give rise to the same statistics in the randomised scenario, we choose to work with the singlet state $|\psi^-\rangle$. The presented proof follows the lines of~\cite{Liang_2010} and extends it to our setting. As noted in the main text we want to find $\mathcal{P}_V$ for the correlation part of (\ref{eq:chsh_eta}) as it would provide us with a lower bound on the violation probability of an entire inequality. For convenience, we rewrite the inequality (\ref{eq:chsh}) with correlation functions. Now, the local bound is $L=2$, and we are interested in observing
\begin{eqnarray}
    I'&=&\eta_A \eta_B |E(\Vec{a}_1,\Vec{b}_1)+E(\Vec{a}_1,\Vec{b}_2)+E(\Vec{a}_2,\Vec{b}_1)\\
    &-&E(\Vec{a}_2,\Vec{b}_2)|+2(1-\eta_A)(1-\eta_B)>2, \nonumber
\end{eqnarray}
as it is equivalent to $\Tilde{Q}>3$ in (\ref{eq:chsh_eta}). Note that the $2(1-\eta_A)(1-\eta_B)$ term vanishes in the asymmetric case but can significantly improve the symmetric scenario results.

\begin{figure}
    \centering
    \includegraphics[scale=0.6]{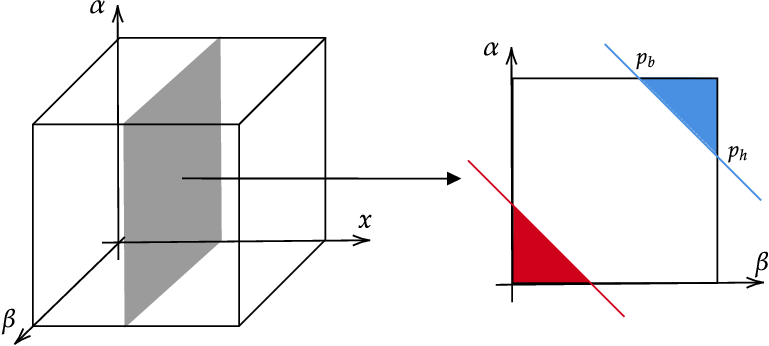}
    \caption{A visual proof help for calculating the lower bound on the probability of Bell inequality violation $\mathcal{P}_V^c$. It shows a three-dimensional cube containing the possible value of the correlation part of the CHSH expression. The red and blue lines for each slice represent the local bound and separate the classically allowed region from the quantum one (red and blue). Calculating the area of each coloured triangle and integrating over $x$ yields a single correlation part of the CHSH inequality violation probability.}
    \label{fig:cube_proof}
\end{figure}

For a given random measurement directions $\Vec{a}_i$ and $\Vec{b}_j$ the correlation function of the examined state is given as $E(\Vec{a}_i,\Vec{b}_j)=-\Vec{a}_i \cdot \Vec{b}_j$. Defining $\vec{r}=\Vec{b}_1+\Vec{b}_2$ and $\vec{r}_{\perp}=\Vec{b}_1-\Vec{b}_2$, expression $I'$ becomes
\begin{eqnarray}
    I'&=&\sqrt{2}\eta_A \eta_B |\sqrt{1+x}\Vec{a}_1 \cdot \hat{r}+\sqrt{1-x}\Vec{a}_2 \cdot \hat{r}_{\perp}|\\
    &+&2(1-\eta_A)(1-\eta_B), \nonumber
\end{eqnarray}
where $\hat{\cdot}$ denotes a unit vector and $x=\Vec{b}_1 \cdot \Vec{b}_2$. The violation we are interested in now is $|\sqrt{1+x}\Vec{a}_1 \cdot \hat{r}+\sqrt{1-x}\Vec{a}_2 \cdot \hat{r}_{\perp}|>[2-2(1-\eta_A)(1-\eta_B)]/\sqrt{2}\eta_a \eta_B$. As argued in~\cite{Liang_2010}, from the rotational symmetry of the problem, directions of the unit vectors are irrelevant and to determine the probability of violation it is enough to sample $\alpha=\Vec{a}_1 \cdot \hat{r}, \beta=\Vec{a}_1 \cdot \hat{r}$, and $x$ uniformly from $[-1,1]$. Geometrically the variables $\alpha, \beta,x$ span a cube with vertices at $\pm 1$ and a volume of $2^3$. Now, we ask what fraction of this cube leads to a violation of $I'<2$. In the positive ($\mathrm{I}$) and negative ($\mathrm{II}$) domain of the expression within the modulus of $I'$, the violation condition defines two linear inequalities
\begin{eqnarray*}
    \mathrm{I:} \, \alpha > \frac{[2-2(1-\eta_A)(1-\eta_B)]/\sqrt{2}\eta_a \eta_B - \beta \sqrt{1+x}}{\sqrt{1-x}} \\
    \mathrm{II:} \,  \alpha < -\frac{[2-2(1-\eta_A)(1-\eta_B)]/\sqrt{2}\eta_a \eta_B + \beta \sqrt{1+x}}{\sqrt{1-x}}.
\end{eqnarray*}
Fixing $x$ we can consider slices of a cube and examine the area of triangles defined by the above equations, see Fig.~\ref{fig:cube_proof}. Since they contribute in an equal manner, we will focus on the region $\mathrm{I}$. The points defining the base and height of the triangle are given as  $p_{b/h}=([2-2(1-\eta_A)(1-\eta_B)]/\sqrt{2}\eta_a \eta_B - \sqrt{1 \mp x})/\sqrt{1 \pm x}$.
From that, the base length and its height are given as $1-p_{b/h}$ respectively and yield the triangle area
\begin{equation*}
    S_{\Delta}=\frac{\left(\sqrt{2} (\eta_B+\eta_A) -\eta_A\eta_B \left(\sqrt{1-x}+\sqrt{x+1}+\sqrt{2}\right)\right)^2}{4 \eta_A^2 \eta_B^2 \sqrt{1-x^2}}
\end{equation*}
The total area of two regions for which the violation occurs is equal to $2S_{\Delta}$. Due to the detection efficiency dependence, we need to include some additional constraints on the possible values of $x$, i.e., the range of $x$ for which the triangle has a non-zero area. This condition is equivalent to $p_{h/b} \in (0, \pm 1)$. From now on, we will focus on two especially relevant cases, but a general reasoning is still possible to perform. These two cases are $\eta_A=1, \eta_B=\eta_{asym}$ and $\eta_A=\eta_B=\eta_{sym}$ referred to as asymmetric and symmetric detection efficiencies in the main text. In the asymmetric case, $x$ has to be contained within $\pm \sqrt{\eta_{asym}^2-1}/\eta_{asym}^2$ which leads to an upper bound on the critical efficiency of $\eta^c_{asym}=1/\sqrt{2}\approx 0.7071$. In the latter the possible values of $x$ are $\pm [\left(\eta_{sym} ^2-2 \eta_{sym} +2\right)$ $ \sqrt{\eta_{sym} ^4+4 \eta_{sym} ^3-8 \eta_{sym} ^2+8 \eta_{sym} -4}]/\eta_{sym} ^4$ and $\eta^c_{sym}=\sqrt{5 + 4 \sqrt{2}}- \sqrt{2} -1\approx 0.8503$. 

To obtain the minimal probability of a single CHSH inequality violation one has to integrate the expression for $2S_{\Delta}$ over $x$ respecting the above bounds and divide it by the cube's volume. Before we do that, note that the two-setting two-outcome scenario is fully characterised by the set of four CHSH inequalities, and thus we can obtain the $\mathcal{P}_V^c$ by multiplying a single inequality violation probability by four. Performing the last two steps, we finally get the lower bounds presented in the main text.

\section{Critical detection efficiency for the GHZ state in a two-setting three-party scenario}
\label{app:ghz}

We present a symmetric Bell inequality for a three-party scenario with $m=2$
and the critical efficiency of $\eta_{\text{crit}}=(\sqrt{10}-1)/3\approx 0.7208$. This threshold is lower than that of the Mermin setup~\cite{Mermin_1990}, which requires \(\eta_{\text{crit}}=0.75\) (provided here for comparison). In both cases, the no-detection events are addressed by introducing an additional third outcome.

\subsection{Inequality}
We consider a Bell scenario with three observers, where each observer can choose between two measurements that yield three possible outcomes. In this case, the probability distribution is given by the set $\{P(a,b,c|x,y,z)\}$, where the measurement settings (inputs) take $x,y,z\in\{0,1\}$ and the outcomes are $a,b,c\in \{+1,-1,0\}$. A general three-party three-outcome Bell inequality can then be written as follows
\begin{equation}
I_{ABC}=\sum_{a,b,c}\sum_{x,y,z=0,1}B_{abcxyz}P(a,b,c|x,y,z)\le 0,
\end{equation}
with a local bound of zero and with the indices $a,b,c$ running over $\{+1,-1,0\}$. However, by imposing the non-signaling conditions and the normalization (i.e., the probabilities sum up to one for each triple of inputs ($x,y,z$)), one may write a generic three-outcome Bell inequality in a form without explicitly including the third outcome ($0$). This is known as the Collins-Gisin format~\cite{Collins_2004}:
\begin{align}
I_{ABC}=&\sum_{a,b,c,x,y,z}\alpha_{abcxyz}P(a,b,c|x,y,z)\nonumber\\
&+\sum_{a,b,x,y}\beta_{abxy}P(a,b|x,y)
+\sum_{a,c,x,z}\beta_{acxz}P(a,c|x,z)
\nonumber\\
&+\sum_{b,c,y,z}\beta_{bcyz}P(b,c|y,z)
+\sum_{a,x}\gamma_{ax}P(a|x)
\nonumber\\
&+\sum_{b,y}\gamma_{by}P(b|y)
+\sum_{c,z}\gamma_{cz}P(c|z)\le 0,
\label{eq:CG3}
\end{align}
where the outcomes are restricted to $a,b,c\in\{\pm 1\}$ and the inputs are $x,y,z\in\{0,1\}$. 

For each triple of measurement settings ($x,y,z$), we define a three-party correlation function and the sum of probabilities as
\begin{align}
&E_{xyz}=\sum_{a,b,c=\pm 1}abcP(a,b,c|x,y,z),\nonumber\\
&O_{xyz}=\sum_{a,b,c=\pm 1}P(a,b,c|x,y,z).
\end{align}
Similarly, the two-party quantities are defined by
\begin{align}
&O_{AB}=\sum_{a,b=\pm 1}\sum_{x,y=0,1}P(a,b|x,y),\nonumber\\
&O_{AC}=\sum_{a,c=\pm 1}\sum_{x,y=0,1}P(a,c|x,z),\nonumber\\
&O_{BC}=\sum_{b,c=\pm 1}\sum_{y,z=0,1}P(b,c|y,z).
\end{align}

Using a linear programming approach, we found a new tight facet-defining three-party Bell inequality, $\langle I_{ABC}^{(1)} \rangle \leq 0$, in the three-outcome probability space. This inequality, as we show below, provides a critical detection efficiency of $\eta_{\text{crit}}\approx 0.7208$ using a rotated version of the GHZ state and orthogonal qubit observables. In particular, we have the expression
\begin{equation}
\label{eq:IABC1} 
I_{ABC}^{(1)}=I_{222}+J_{222}-K_{22}.
\end{equation}
This Bell inequality $\langle I_{ABC}^{(1)} \rangle \leq 0$ is a special class of inequalities~(\ref{eq:CG3}) with zero single-party marginal coefficients. The three separate terms that build up the expression are defined by
\begin{align}
\label{eq:Icomp1}
I_{222}& = 3E_{000}-{\rm sym}[E_{001}]-3{\rm sym}[E_{011}]+E_{111},\nonumber\\
J_{222}& =-O_{000}+2{\rm sym}[O_{001}]-{\rm sym}[O_{011}]+2O_{111},\nonumber\\
K_{22}& = O_{AB}+O_{AC}+O_{BC},
\end{align}
where we use here a compact notation 
\begin{eqnarray}
{\rm sym}[E_{klm}] = \sum_{\pi(k,l,m)} E_{klm},
\end{eqnarray}
for symmetrizing over different observers, with the sum taken over all permutations $\pi(k,l,m)$ of the indices $(k,l,m)$. For example, ${\rm sym}[E_{011}] = E_{011} + E_{101} + E_{110}$, which corresponds to the permutations of $k=0,l=1,m=1$. 

We also provide a Mermin-type Bell expression, $\langle I_{ABC}^{(2)} \rangle \leq 0$, which leads to a critical detection efficiency of $\eta_{\text{crit}}=0.75$. The Bell expression is given by
\begin{equation}
\label{eq:IABC2} 
I_{ABC}^{(2)}=\tilde{I}_{222}+\tilde{J}_{222}-\tilde{K}_{22},
\end{equation}
where 
\begin{align}
\label{eq:Icomp2}
\tilde{I}_{222}& = E_{000}-{\rm sym}[E_{011}],\nonumber\\
\tilde{J}_{222}& = O_{111}+{\rm sym}[O_{001}],\nonumber\\
\tilde{K}_{22}& = (O_{AB}+O_{AC}+O_{BC})/2.
\end{align}
Notice that $\tilde{I}_{222}$ is a Mermin-type term~\cite{Mermin_1990}, now defined in the space of three-outcome probability distributions. 
    
\subsection{Critical detection efficiency}
Now, we compute $\eta_{\text{crit}}$ corresponding to a quantum distribution generated using a rotated GHZ state and orthogonal observables in the Bell inequality (\ref{eq:IABC1}). Note that the state below is equivalent to the original GHZ state under local unitaries and they are indistinguishable in the random measurements setting. The state under consideration is given by
\begin{equation}
\label{eq:rGHZ}
|GHZ(\varphi)\rangle=\frac{|000\rangle+e^{i\varphi}|111\rangle)}{\sqrt 2},    \end{equation}
where we fix $\varphi=\arctan{(1/3)}$, and every observer measures the same set of two observables:
\begin{equation}
\label{eq:obs}
A_0=B_0=C_0=\sigma_x\, \text{ and }\, A_1=B_1=C_1=\sigma_y. 
\end{equation}
The three-party quantum correlations $E^Q_{xyz}$ are then given by
\begin{align}
\label{eq:EQxyz}
E^Q_{000} &= \frac{3}{\sqrt{10}},\quad &E^Q_{011}=E^Q_{101}=E^Q_{110} &= -\frac{3}{\sqrt{10}},\nonumber\\[1mm]
E^Q_{111} &= \frac{1}{\sqrt{10}},\quad &E^Q_{100}=E^Q_{010}=E^Q_{001} &= -\frac{1}{\sqrt{10}}.
\end{align}
For perfect detectors ($\eta=1$) the third outcome never occurs, so that 
\begin{equation}
\label{eq:OQxyz}
O^Q_{x,y,z}=1\quad \text{for all } x,y,z\in\{0,1\}.
\end{equation}
Furthermore, from the properties of the rotated GHZ state~(\ref{eq:rGHZ}) and the fact that our observables are $\pm 1$-valued, one finds
\begin{equation}
\label{eq:OQxy}
O^Q_{AB}=O^Q_{AC}=O^Q_{BC}=16\times\frac{1}{4}=4.    
\end{equation}
By inserting these values into~Eqs.~(\ref{eq:EQxyz}),(\ref{eq:OQxyz}), and (\ref{eq:OQxy}), we obtain 
\begin{equation}
I^Q_{222}=\sqrt{160},\quad J^Q_{222}=4,\quad \text {and}\quad K^Q_{22}=12.    
\end{equation}

Next, let us consider the case of finite efficiency of the detectors. In the symmetric scenario, i.e. $\eta_A=\eta_B=\eta_C=\eta$, the $\eta$-dependent quantities become
\begin{align}
&E_{xyz}^Q(\eta)=\eta^3 E_{xyz}^Q, \nonumber\\
&O_{xyz}^Q(\eta)=\eta^3O_{xyz}^Q    
\end{align}
for all $x,y,z\in\{0,1\}$ and 
\begin{align}
&O^Q_{AB}(\eta)=\eta^2O_{AB}^Q, \nonumber\\
&O^Q_{AC}(\eta)=\eta^2O_{AC}^Q, \nonumber\\
&O^Q_{BC}(\eta)=\eta^2O_{BC}^Q.     
\end{align}
Plugging these values into Eq.~(\ref{eq:IABC1}) yields the quantum value as a function of $\eta$:
\begin{equation}
\label{eq:IABCeta}
    I^{(1)}_{Q,ABC}(\eta)=\eta^3 I^Q_{222}+\eta^3 J^Q_{222} +\eta^2K^Q_{22}.
\end{equation}
\\
For a non-zero quantum violation, this expression should exceed the local bound of zero. Hence, the critical detection efficiency $\eta_{\text{crit}}$ is given by the condition $I^{(1)}_{Q,ABC}(\eta)=0$. This leads to  
\begin{equation}
\label{eq:etacrit1}
    \eta^{(1)}_{\text{crit}} = \frac{K_{22}^Q}{I^Q_{222}+J^Q_{222}}=\frac{\sqrt{10}-1}{3}\approx 0.7208
\end{equation}
for the Bell expression~(\ref{eq:IABC1}). 

Concerning the Mermin-type expression~(\ref{eq:IABC2}), if we use the state~(\ref{eq:rGHZ}) with $\varphi = 0$ and the observables given by Eq.~(\ref{eq:obs}), we have  
\begin{equation}
\tilde{I}^Q_{222}= 4,\quad \tilde{J}^Q_{222}=4,\quad \tilde{K}^Q_{22}=6.    
\end{equation}
Thus, the corresponding critical detection efficiency is 
\begin{equation}
\label{eq:etacrit2}
    \eta^{(2)}_{\text{crit}} = \frac{\tilde{K}_{22}^Q}{\tilde{I}^Q_{222}+\tilde{J}^Q_{222}}=0.75,
\end{equation}
This result has been proved to be optimal for the Mermin setup~\cite{Larsson_1998}.

\section{Probability of violation data tables}
\label{app:table}
\begin{table}
    \centering
    \caption{Relative growth in the probability of violation for the maximally entangled two-qubit state and symmetric detection efficiency $\eta$ when increasing the number of random measurements from $m^A=m^B=m_1$ to $m_2$.} \label{tab:growth}
    \label{tab:prob_change}
    \begin{tabular}{|c|c|c|c|c|}
    \hline
     & \multicolumn{4}{c|}{$\Delta \mathcal{P}_{m_1  \rightarrow m_2}$ [\%]} \\
    \hline
    $\eta$ & $2 \rightarrow 3$ & $3 \rightarrow 4$ &$4 \rightarrow 5$ & $5 \rightarrow 6$ \\
     \hline
    $0.83$ & $1404.25$ & $429.41$ & $159.26$ & $542.86$ \\
    \hline
    $0.84$ & $775.00$ & $309.15$ & $178.23$ & $142.75$ \\
    \hline
    $0.85$ & $742.06$ & $290.52$ & $161.59$ & $99.66$ \\
    \hline
    $0.86$ & $695.78$ & $256.83$ & $129.67$ & $62.76$ \\
    \hline
    $0.87$ & $652.37$ & $219.14$ & $96.83$ & $41.46$ \\
    \hline
    $0.88$ & $602.53$ & $187.41$ & $73.31$ & $28.22$ \\
    \hline
    $0.89$ & $555.17$ & $158.45$ & $54.57$ & $20.87$ \\
    \hline
    $0.90$ & $507.93$ & $133.35$ & $43.33$ & $13.49$ \\
    \hline
    $0.91$ & $463.27$ & $111.80$ & $32.94$ & $9.71$ \\
    \hline
    $0.92$ & $421.10$ & $93.18$ & $25.85$ & $7.21$ \\
    \hline
    $0.93$ & $381.43$ & $78.88$ & $19.86$ & $4.95$ \\
    \hline
    $0.94$ & $346.40$ & $65.41$ & $15.65$ & $3.51$ \\
    \hline
    $0.95$ & $311.13$ & $55.49$ & $12.06$ & $2.34$ \\
    \hline
    $0.96$ & $280.05$ & $46.27$ & $9.25$ & $1.93$ \\
    \hline
    $0.97$ & $250.78$ & $38.85$ & $7.14$ & $1.38$ \\
    \hline
    $0.98$ & $224.10$ & $32.12$ & $6.00$ & $0.88$ \\
    \hline
    $0.99$ & $199.02$ & $27.32$ & $4.41$ & $0.59$ \\
    \hline
    $1.00$ & $176.23$ & $22.95$ & $3.47$ & $0.45$ \\
    \hline
    \end{tabular}
\end{table}
In Table~\ref{tab:growth}, we provide the data concerning the relative growth in the probability of violation for the maximally entangled two-qubit state and symmetric detection efficiency when increasing the number of random measurements from $m^A=m^B=m_1$ to $m_2$. For a more detailed discussion, see Sec.~\ref{sec:typicality}.

\bibliographystyle{apsrev4-2}
\bibliography{bibliography}

\end{document}